\documentclass[useAMS,usenatbib,fleqn]{mnras}
\usepackage{textcomp}
\usepackage{amsmath}
\usepackage{amsfonts}
\usepackage{amssymb}
\usepackage{graphicx}
\usepackage[usenames, dvipsnames]{color}

\def \apj {ApJ}

\def \apjl {ApJL}
\def \apjs {ApJS}
\def \aap {A\&A}

\def \araa {Annu. Rev. Astro. Astrophys.}

\def \mnras {MNRAS}

\begin{document}

\date{Preparation: december 2018}

\title[3D magnetar outbursts]{Triggering magnetar outbursts in 3D force-free simulations} 

\author[F. Carrasco et al.]{
	Federico Carrasco,$^{1}$\thanks{E-mail: federico.carrasco@uib.es}
	Daniele Vigan\`o,$^{1}$
	Carlos Palenzuela,$^{1}$
	Jose A. Pons$^{2}$
	\\
	% List of institutions
	$^{1}$Departament de F\'i{}sica \& IAC3, Universitat de les Illes Balears and Institut d'Estudis Espacials de Catalunya, Baleares E-07122, Spain. \\ %Palma de Mallorca,
	$^{2}$Departament de F\'{\i}sica Aplicada, Universitat d'Alacant, Ap. Correus 99, 03080 Alacant, Spain.
}

\label{firstpage}
\pagerange{\pageref{firstpage}--\pageref{lastpage}}
\maketitle

\begin{abstract}

In this letter, we present the first 3D force-free general relativity simulations of the magnetosphere dynamics related to the magnetar outburst/flare phenomenology. 
Starting from an initial dipole configuration, we adiabatically increase the helicity by twisting the footprints of a spot on the stellar surface and follow the succession of quasi-equilibrium states until a critical twist is reached.
Twisting beyond that point triggers instabilities that results in the rapid expansion of magnetic field lines, followed by reconnection, as observed in previous axi-symmetric simulations. 
If the injection of magnetic helicity goes on, the process is recurrent, periodically releasing a similar amount of energy, of the order of a few \% of the total magnetic energy.
From our current distribution, we estimate the local temperature assuming that dissipation occurs mainly in the highly resistive outermost layer of the neutron star.
We find that the temperature smoothly increases with injected twist, being larger for spots located in the tropical regions than in polar regions, and rather independent of their sizes.
After the injection of helicity ceases, the magnetosphere relaxes to a new stable state, in which the persistent currents maintain the footprints area slightly hotter than before the onset of the instability.

% of the reference value, assumed to be the total energy content in the dipole configuration, $W_0 \sim  B_{15}^2 R_{10}^3 \times 10^{47}$ erg. 

\end{abstract}


%%%%%%%%%%%%%%%%%%%%%%%%%%%%%%%%%%%%%%%%%%%%%%%%%%%%%%%%%%%%%%%%%%%%%%%%%%%%%%%%%%%%%%%%%%%%%%%%%%%%%%%%%%%%%%%%%%%%%%%%%%%%%%%%%%%
\section{Introduction}
%\noindent{\bf{\em Introduction.}}
%%%%%%%%%%%%%%%%%%%%%%%%%%%%%%%%%%%%%%%%%%%%%%%%%%%%%%%%%%%%%%%%%%%%%%%%%%%%%%%%%%%%%%%%%%%%%%%%%%%%%%%%%%%%%%%%%%%%%%%%%%%%%%%%%%%

Magnetars \cite{olausen14,kaspi17,mereghetti15}, arguably the most magnetized stable objects in the universe, are relatively slowly rotating neutron stars with spin periods of $P\sim 1-10$ s and estimated magnetic fields of  $B\sim 10^{14}-10^{15}$ G. The number of the known magnetars has been steadily increasing \cite{olausen14}\footnote{See the McGill catalog: {\tt http://www.physics.mcgill.ca/$\sim$ pulsar/magnetar/main.html}} since the mid 90's where the main aspects of the magnetar theory were proposed \cite{duncan92,thompson93,thompson95,thompson96}.
Its most relevant distinguishing feature, compared to common radio-pulsars, is that the observed electromagnetic luminosity is well above the loss of rotational energy, indicating that the magnetic field is the energy source, instead of rotation. Magnetars are also characterized by energetic transient phenomena, detected in X-rays and $\gamma-$rays, ranging from short bursts lasting fractions of second, to long outbursts during which their luminosity suddenly increases, slowly returning to a low-state on long timescales (months or even years) \cite{Rea2011,coti18}, and sporadic giant flares (observed in three cases). All these events are supposed to be linked to reconnection events of the magnetospheric magnetic field, possibly triggered by the star's interior evolution.
Thus, understanding the cause and the dynamics of magnetospheric activity becomes a key issue in magnetar theory.

In the rarefied magnetospheric environment, the electromagnetic forces dominate over particle inertia and allows to assume the force-free approximation, which has been widely used to study global properties of pulsars~\cite{gruzinov99, contopoulos1999, mckinney2006relativistic, timokhin2006force, spitkovsky2006}.
This approximation has also been adopted to model magnetar outbursts,  studying the dynamical response of the magnetosphere under the injection of magnetic helicity from a localized spot on the surface. If the injection timescale (determined by the internal field evolution) is much longer than the rapid response of the magnetosphere (milliseconds), 
the evolution of the twisted magnetic configurations can be modeled as an adiabatic sequence of quasi-stationary states, at least as long as the configurations are stable.
Once a magnetic line bundle is twisted above a critical value, the system becomes unstable and eventually undergoes a large scale magnetic reconnection and plasmoid emission. 
The dynamics of these events has been extensively studied by analytical and numerical models considering axi-symmetric configurations~\cite{beloborodov07,beloborodov2009, parfrey2013, pili2015, chen2016}.

In this work we aim at extending previous models in two different directions. First, we include General Relativity (GR) effects by considering the Schwarzschild metric, as a good approximation to describe the spacetime surrounding a relatively slow rotating star. 
Second, and more important, by performing for the first time 3D simulations without any symmetry restriction. %This point is crucial, since the dynamics of the twisted magnetic spots, which a priori does not have any symmetry, can lead to remarkable differences with respect to the 2D case.

%%%%%%%%%%%%%%%%%%%%%%%%%%%%%%%%%%%%%%%%%%%%%%%%%%%%%%%%%%%%%%%%%%%%%%%%%%%%%%%%%%%%%%%%%%%%%%%%%%%%%%%%%%%%%%%%%%%%%%%%%%%%%%%%%%%%% 
\section{ Numerical Setup}
%\noindent{\bf{\em Numerical Setup.}}
%%%%%%%%%%%%%%%%%%%%%%%%%%%%%%%%%%%%%%%%%%%%%%%%%%%%%%%%%%%%%%%%%%%%%%%%%%%%%%%%%%%%%%%%%%%%%%%%%%%%%%%%%%%%%%%%%%%%%%%%%%%%%%%%%%%%%

We use the version of the GR force-free electrodynamics formalism derived in \cite{FFE}, which is well-posed and involves the full force-free current density. %~\footnote{Similar hyperbolic formulations were presented in Refs.~\cite{pfeiffer,Pfeiffer2015}.}. 
More concretely, we implement the evolution system given by Eqs.~(8)-(10) in \cite{NS}.
The numerical scheme to solve these equations is based on the \textit{multi-block approach} \cite{Leco_1, Carpenter1994, Carpenter1999, Carpenter2001, Leco_2}. 
We employ difference operators which are eight-order accurate on the interior and fourth-order at the boundaries, with adapted Kreiss-Oliger dissipation operators \cite{Tiglio2007}. 
A classical fourth order Runge-Kutta algorithm is used for time integration. Our numerical grid extends from an interior sphere at radius $r=R$, that represents the star radius, to an exterior spherical surface at $r\sim 50 R$.
%This region is covered by a total of $10\times 6$ grids, being $10$ the number of layers. The grids layers do not cover regions of identical radial extension, having more resolution near the inner boundary than in the asymptotic region: from layer to layer we decrease the effective radial resolution by a factor $1.35$. 
We adopt a resolution $N_{\theta} \times N_{\phi} \times N_{r}$ with $N_{\phi}=321$, $N_{\theta} = 161$, while $N_{r}$ is adjusted to satisfy $\Delta r \lesssim r \Delta \theta$ everywhere in the domain.

The stellar surface is assumed to behave as an idealized perfect conductor. Thus, at the surface, the normal component of the magnetic field is fixed
and the electric field is prescribed from the ideal MHD condition consistent with the shear perturbations at the stellar crust.
At the outer boundary, on the other hand, we set non-reflective conditions to allow all perturbations to propagate away. % and no spurious noises to arise.
The numerical implementation of such boundary conditions and how to handle current sheets have been detailed in~\cite{NS, FFE2}.
%
%In order to handle current sheets, we use a rather standard approach in which electric field is effectively dissipated to maintain the condition that the plasma is magnetically dominated (i.e.~$B^2 -E^2 >0$). 

We consider an initial magnetic dipolar field configuration %with dipole moment $\mu$, %which can be written as an exact vacuum solution in a Schwarzschild 
%space-time \cite{Shapiro1983}:
%\begin{equation}
% A_{\phi} = \frac{3\mu \sin^2 (\theta)}{4M} \left[ 1 + \frac{r^2}{2M^2} \ln (1-2M/r) +  \frac{r}{M} \right] \label{A-potential}
%\end{equation}
%where $\mu$ is the dipole moment and $M$ the stellar mass.
in a Schwarzschild external region $r>R$, of mass $M$ \cite{Shapiro1983}. 
We adopt geometrized units in which $c=G=1$, and Lorentz-Heaviside units for the electromagnetic field. 
A typical neutron star of compactness $M/R = 0.2$ is considered. 
In our coordinates, the magnetic moment forms an angle $\chi$ with the $z$-axis, which represents the center of the circular surface region of the sheared footprints. This misalignment angle $\chi$ is therefore the angle between the NS's magnetic dipole and the center of the twisted region.
The profile describing the perturbation is given by: %$\Omega(\theta)$, like the polar cap studied on Ref.~\cite{parfrey2013}, i.e.:
 \begin{equation}\label{eq:pc}
 \Omega(\theta, t) = \frac{\omega(t)}{1+\exp \left[\kappa (\theta - \theta_{s}) \right] } 
 \end{equation}
where $\theta_{s}$ is the angular extension of the spot, and we set $\kappa=30$.
The value of $\omega(t)$ smoothly increases from zero, at $t=0$, to its final value $\omega_o = 2\times 10^{-3} \left[1/M\right]$, at $t=400 M$. %by using a cosine bell time function.
%and hereafter we set $\omega_o = 2\times 10^{-3} \left[1/M\right]$ as the twist rate at the center of the spot The perturbation is gradually turned at $t=0$, reaching the value of eq.~(\ref{eq:pc}) at $t=400 M$. 

We quantify the accumulated injected twist as, $\psi_{\rm inj}(t) = \int_0^t \omega(t) dt$. The magnetic energy normalized to the initial potential state, $W/W_0$, serves as a measure of the additional magnetic energy stored in the magnetosphere.
We also monitor the currents during the dynamics, which will allow later to estimate the temperature of the surface. 

%\begin{equation}
% \sigma T_{\text{eff}}^4 = \frac{4\pi \eta}{c^2} \Delta r J^2
%\end{equation}
%where $\sigma \simeq 5.67 \times 10^{-5} \, g \, s^{-3} \, K^{-4}$ is the Stefan-Boltzmann constant, $\eta$ is the magnetic diffusivity and $\Delta r$ represents the thickness of the region assumed to radiate as a blackbody, just below the stellar surface . Finally, we express the effective surface temperature around typical values:
%\begin{equation}
% T_{\text{eff}} \approx 0.9 \, keV  \left[ \frac{\Delta r}{1 m} \right]^{1/4} \left[ \frac{\eta}{10^3 cm^2 s^{-1}} \right]^{1/4} \left[ \frac{B}{10^{14} G} \right]^{1/2} \left[ \frac{10 km}{R} \right]^{1/2} \hat{J}^{1/2}
%\end{equation}

%%%%%%%%%%%%%%%%%%%%%%%%%%%%%%%%%%%%%%%%%%%%%%%%%%%%%%%%%%%%%%%%%%%%%%%%%%%%%%%%%%%%%%%%%%%%%%%%%%%%%%%%%%%%%%%%%%%%% RESULTS
\section{Results}
%\noindent{\bf{\em Results.}}

We first focus on a representative case, with a misalignment angle $\chi = 0.15 \pi$ and  angular cap size $\theta_s = 0.15 \pi$, for which the injection is smoothly turned off after some time. %, reaching a final injected twist $\psi_{\rm inj}=3.8$ rad.  
The shearing of the footprints launches Alv\'{e}n waves along the magnetic filed lines, which are reflected back and forth within the closed region of the magnetosphere.  %along magnetic field lines
The complicated pattern of wave-fronts eventually reaches a quasi-equilibrium state, provided that the shearing timescale is much longer than the wave-crossing times of the twisted magnetic lines. As a consequence, the following description dynamics are valid for a large range of $\omega_o$.

\begin{figure}
	\begin{center}
		\includegraphics[scale=0.20]{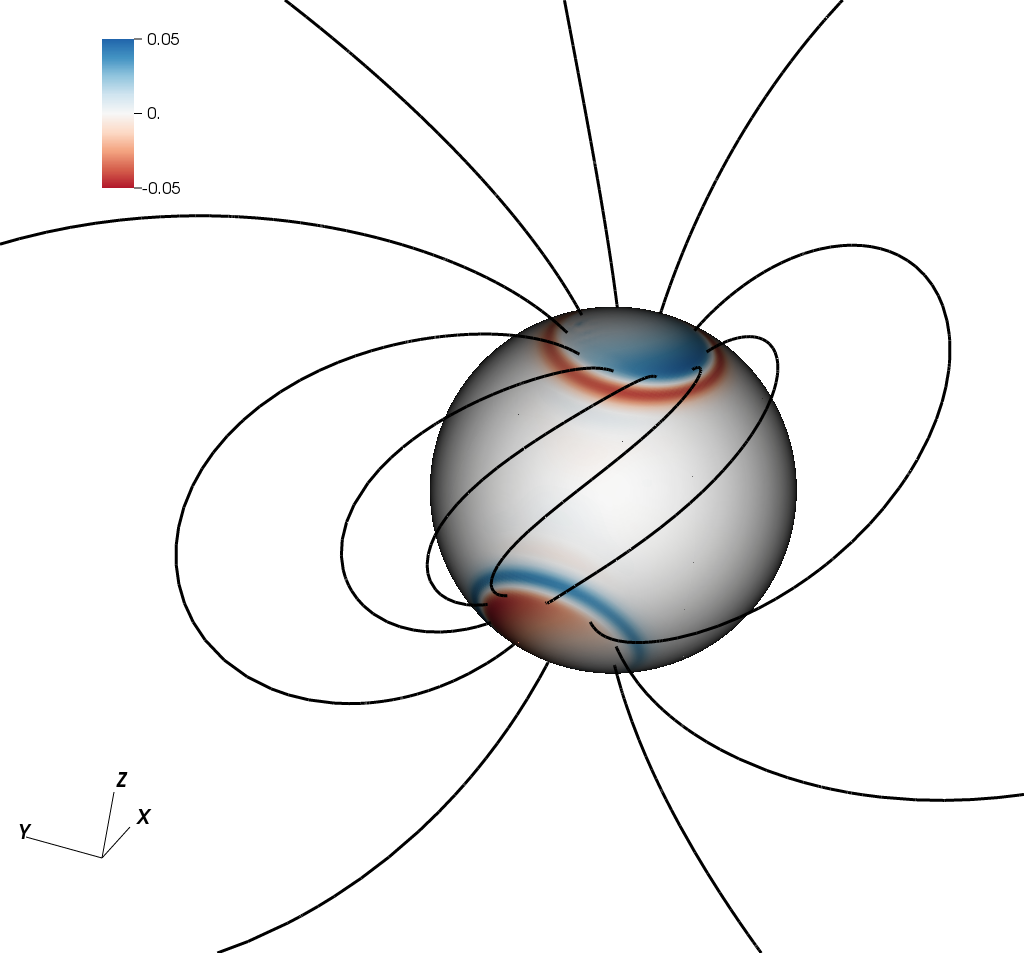}
		\includegraphics[scale=0.20]{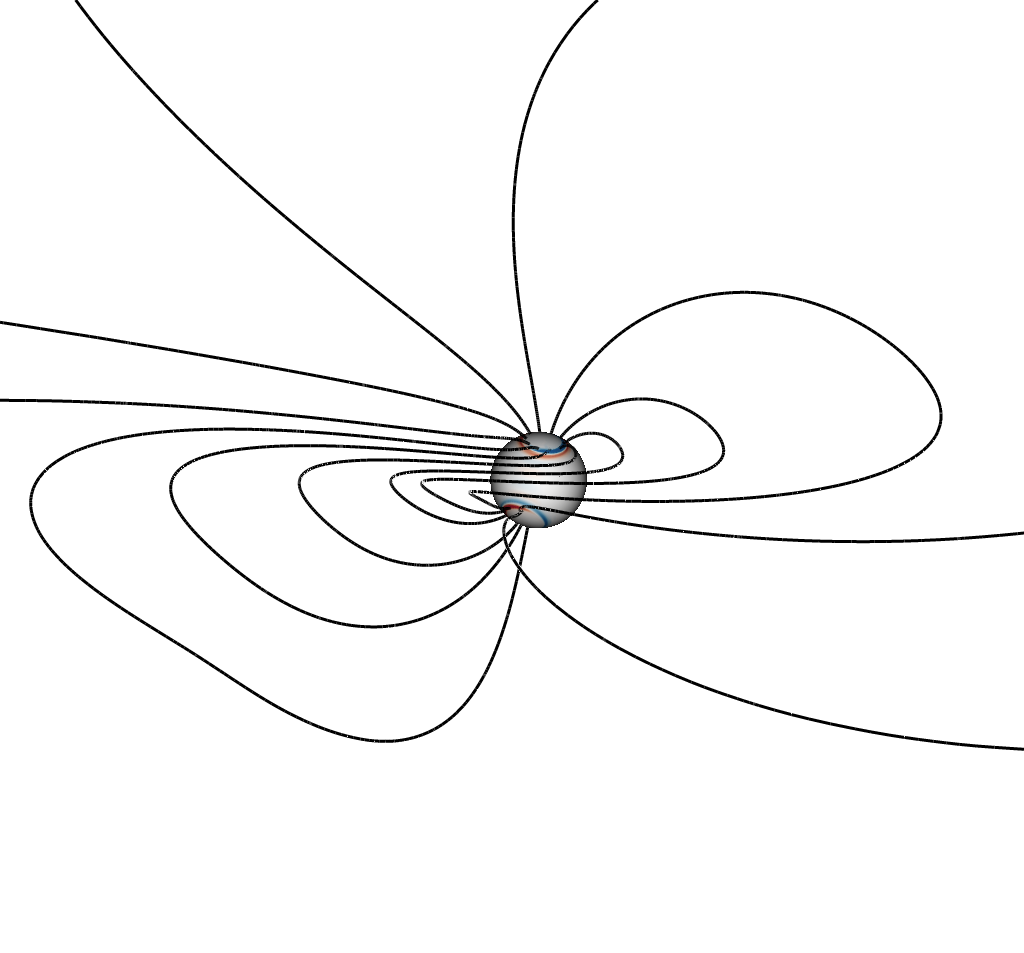}
		\includegraphics[scale=0.20]{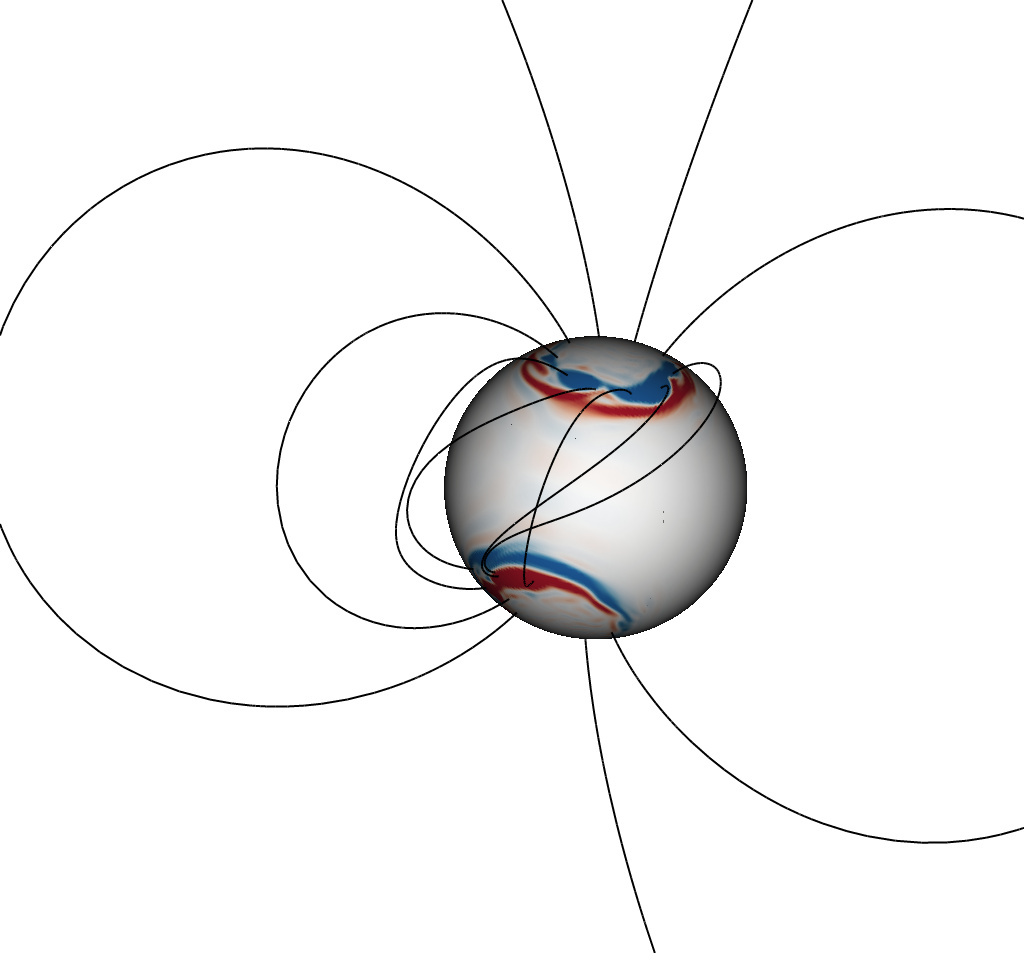}
		\caption{Representative magnetic field lines for rotating spot of size $\theta_s = 0.15\pi$ and misalignment $\chi=0.15\pi$, together with the electric radial current, $J_r$, at the surface, indicated by the color scale. Top: quasi-adiabatic stage, with the magnetic fields moving through quasi-equilibrium configurations. Middle: unstable phase, with an expansion of the magnetic loops, followed by a fast reconnection event. Bottom: relaxed final state, different from the quasi-adiabatic ones. }
		\label{fig:3D}  
	\end{center}
\end{figure}

\begin{figure*}%[htp]
	\centering
\includegraphics[scale=0.15]{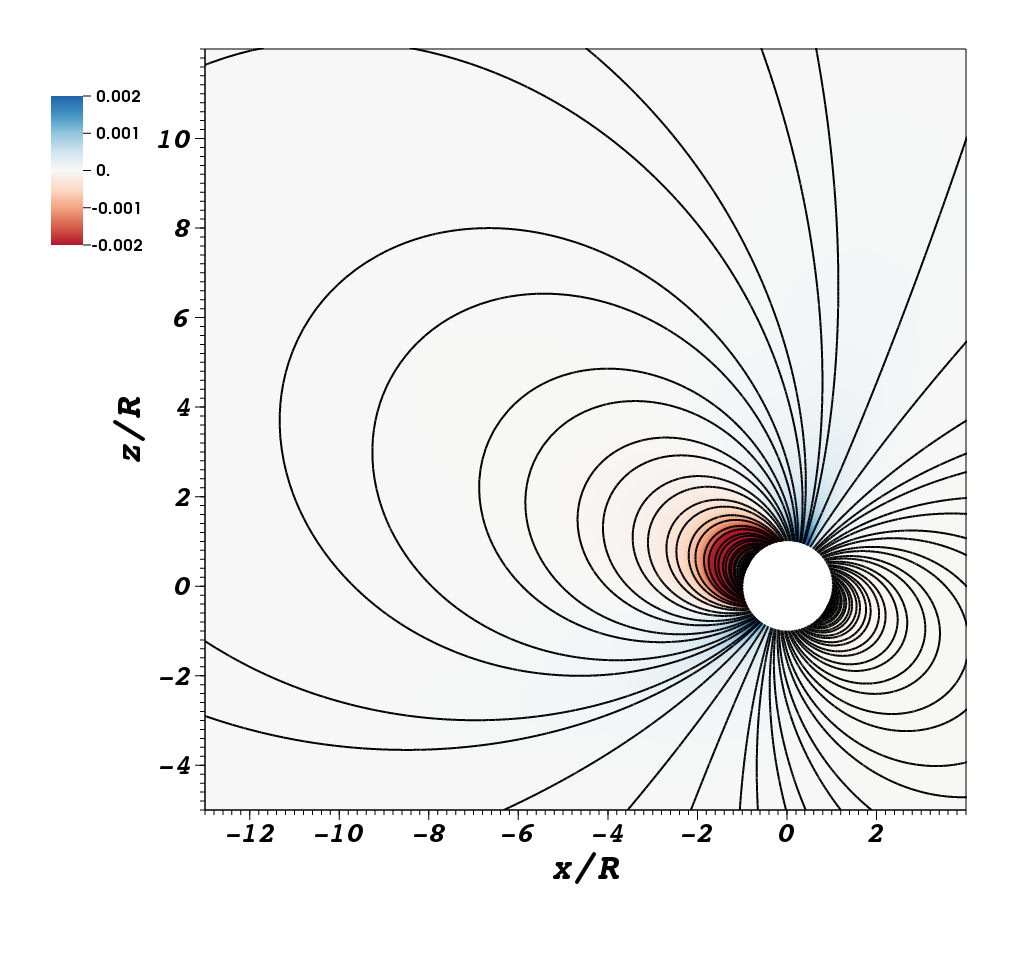}
\includegraphics[scale=0.15]{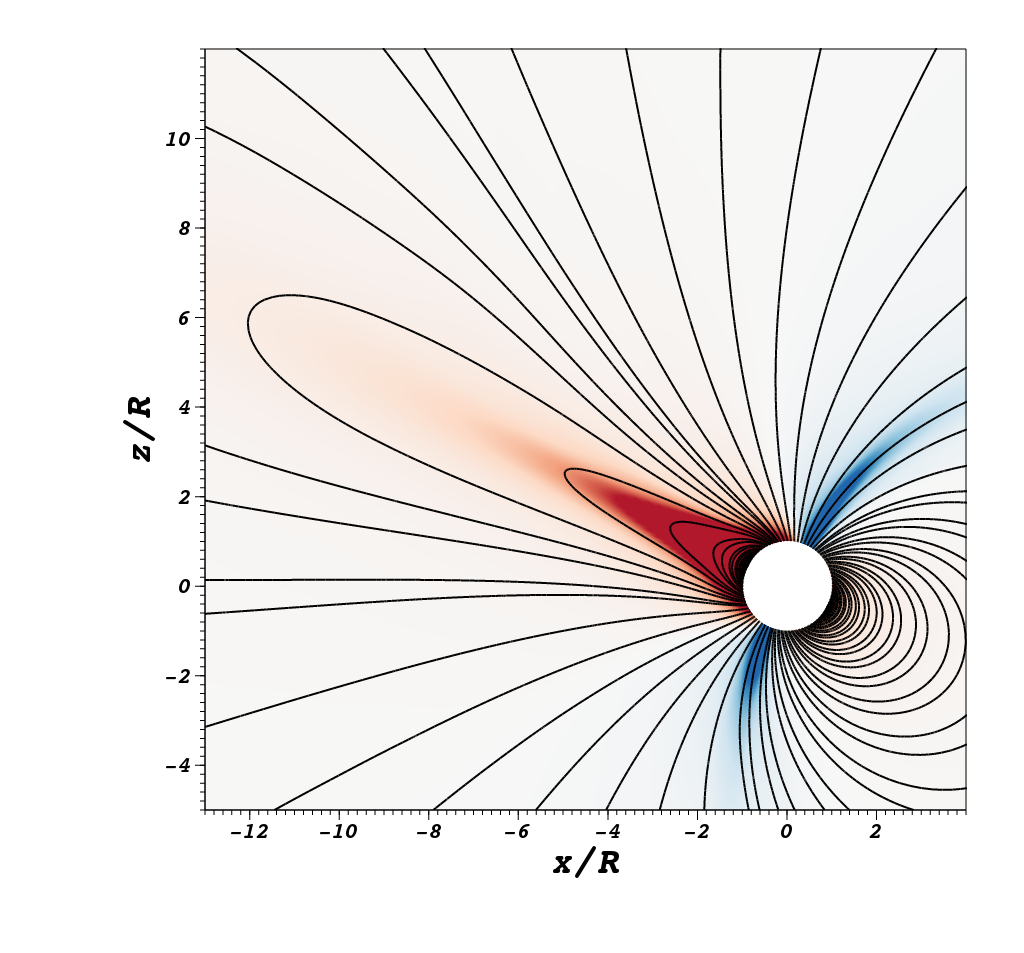}
\includegraphics[scale=0.15]{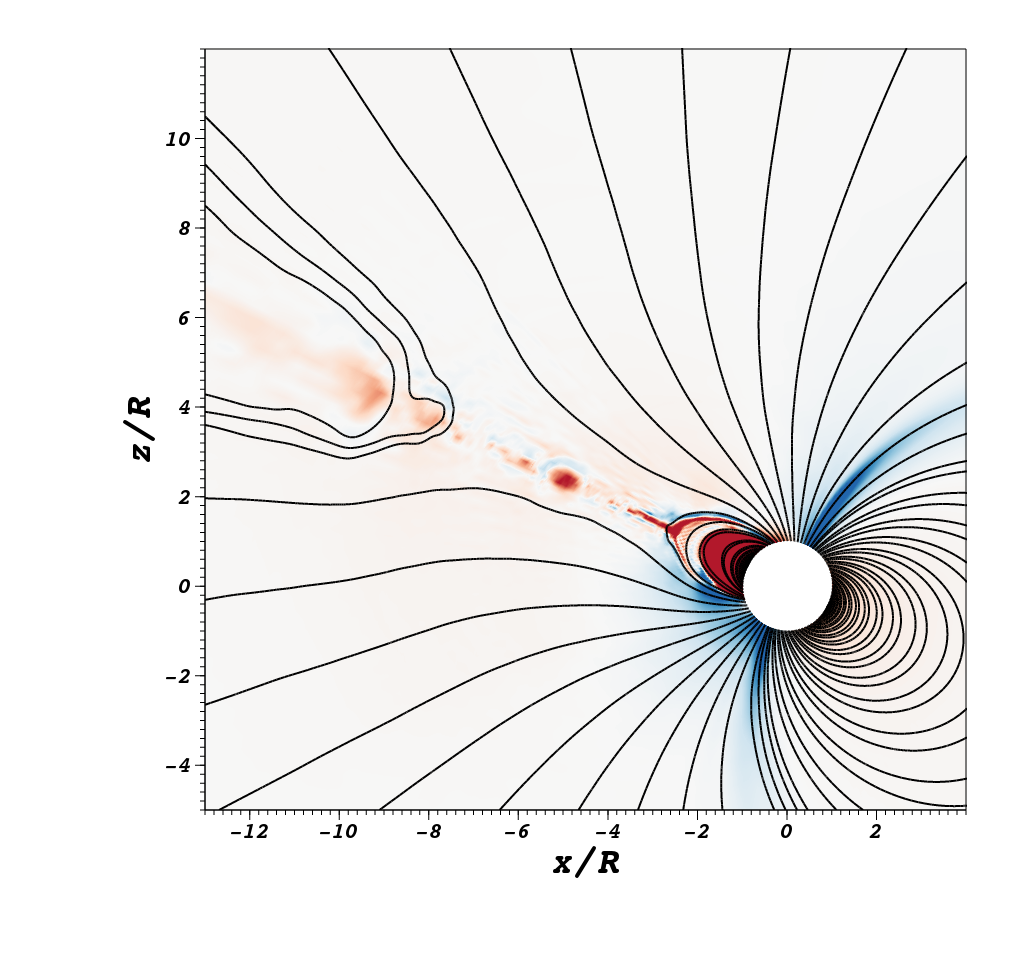}	
	\caption{\textit{Magnetic field evolution for the rotating spot of size $\theta_s = 0.15\pi$ and misalignment $\chi=0.15\pi$.} 
		Lines represent the magnetic field projected in the $x-z$ plane, while the off-plane component is shown in color scale. 
	Left: quasi-equilibrium (non-potential) state. Middle: unstable expansion phase. Right: magnetic reconnection, ejecting plasmoids of different sizes.
	}
	\label{fig:2Dsequence}
\end{figure*}

The qualitative behavior of the magnetospheric response can be described by several stages, illustrated in Fig.~\ref{fig:3D}, where some representative magnetic field lines and the radial component of the current $J^r$ at the stellar surface are shown.
After the first short initial transient described above, the injected helicity gradually modifies the magnetic configuration in an adiabatic succession of non-potential equilibrium states (top panel). 
The evolution of the stored magnetic energy, $W(\psi_{\rm inj})$, and the associated currents, is independent of the injection rate $\omega_o$ (provided it is not too fast).
Beyond a critical  twist, $\psi_{\rm crt} \sim 3$--$4$, the system becomes unstable and departs from the quasi-equilibrium configurations. In this stage (middle panel), some magnetic field lines quickly expand (approaching an open state) and produce a tangential discontinuity in the magnetic field near the equatorial plane~\footnote{
Such critical behavior, in which the field experiences an explosive expansion due to the loss of ideal magnetic equilibrium, has been studied in more detail (in axial symmetry) at e.g.~\cite{mikic1994, uzdensky2002, parfrey2013}.}.
Then, the plasma resistivity (physical in nature, numerical in our case) leads to magnetic reconnection in these current sheets and the subsequent expulsion of plasmoids.
Finally, after the reconnection, the system relaxes into a quasi-equilibrium configuration which differs from those reached during the first quasi-adiabatic stage (bottom panel). Part of the twist and the associated currents remain stable in a rearranged configuration, with a larger magnetic energy than the initial vacuum dipole. 

These sudden reconnection events, release and dissipate significant fractions of the energy stored in the magnetosphere. 
Figure \ref{fig:2Dsequence} illustrates the magnetic field dynamics in a sequence of one such events: the quasi-equilibrium configurations followed by the unstable expansion of the magnetic field, the formation of a current sheet in the equatorial plane, and the subsequent ejection of plasmoids with different sizes. The main feature introduced by the misalignment being that reconnection and emissions are no longer axially-symmetric as for the case $\chi=0$.
There is now a preferred direction, defined by the intersection of the $x-z$ plane with the dipole's equator, along which these plasmoids propagate at nearly the speed of light.
We turn to a detailed analysis of the energetics of these events, potentially available to power outbursts and flares. The accumulated energy prior to reconnection, due to the presence of twisted magnetic fields, constitutes about $2.7 \%$ of the initial energy of the dipole, which can be estimated as $W_0 \sim  B_{15}^2 R_{10}^3 \times 10^{47}$ erg, 
being $B_{15} \equiv \frac{B}{10^{15} {\rm G}}$ and $R_{10} \equiv \frac{R}{10 \rm{km}}$  the normalized values (assumed to be 1 hereafter) of the dipolar magnetic field at the polar surface and of the star's radius. From this available energy reservoir,
approximately $20\%$ is expelled from the system, while nearly $30\%$ is dissipated~\footnote{We compute this quantity by tracking the energy associated with the electric field removed from the system at the current sheet, in order to maintain the condition $B^2 - E^2 >0$, as in other force-free studies. The numerical dissipation is instead negligible.} during the process.
%It clearly depends on the numerical prescription used (also on resolution) and is not accounting for the physical processes involved.
%But it provides, nevertheless, a rough estimation of the total amount of energy converted into heat or radiated away at these regions.}.
%\footnote{Here the dissipation is considered as the energy associated with the electric field numerically reduced in order to satisfy the causality constraint $E<B$, as in other FFE works. The numerical dissipation is instead negligible.}

%Interestingly, this amount of energy (i.e. $\sim B_{15}^2 R_{10}^3 \times 10^{45}$ erg), could be enough to power a giant flare \DV{Tengo duda si mencionar esto arriba.. Habria que decir que el radio deberia ser mayor, para llegar a los 1e47 erg observados en un caso... Me parece un poco debil como resultado. En todo caso lo mas probable es que un giant flare sea una reorganizacion total de la magnetosfera, no solo de un twist. Lo quitamos?}. 

The plasmoids are presumably heated at their formation in the current sheet and propagate both outwards and inwards. 
Those ``plasma fireballs'' trapped by the NS magnetic field were suggested to explain the few minutes tail after the main flash in giant flares, which is modulated by the spin period of the NS. 
On the other hand, part of the energy dissipated at the current sheet might be radiated right away at the explosive reconnection, producing the short $0.1-0.2 \,s $ flash in luminosity \citep[e.g.][]{gourgouliatos2018strongly}. 
The remaining $50\%$ of the energy stored during the injection remains in the magnetosphere, such that the new equilibrium attained after the event is not a potential solution, but a different twisted magnetosphere configuration.
%: only a fraction of the twisted magnetic lines has reconnected and the field rearranges differently. 

Following \cite{akgun2018crust}, we assume that, to close the circuit, part of the magnetospheric currents must be dissipated as they go through the very thin and resistive outer layer of the star ($\Delta r \sim $ few m), with electrical conductivity\footnote{Calculated with the public routines  available at: {\tt http://www.ioffe.ru/astro/conduct/}} $\sigma_e$,  (see \cite{Potekhin} and references therein). 
We can then obtain the map of the estimated surface temperature by equating the local Joule dissipation rate to the blackbody emission from the heated surface: $\sigma T_{\text{bb}}^4 = \Delta r J^2/\sigma_e$, where $\sigma$ is the Stefan-Boltzmann constant, which gives: %{\bf the temperature as seen by an observer at infinity}
%\begin{equation}
%T^\infty_{\text{bb}} \approx  0.18 ~{\rm keV} ~e^\nu \left[ \frac{\Delta r}{1~ {\rm m}}\right]^{\frac{1}{4}} \left[  \frac{10^{17} ~{\rm s}^{-1}}{\sigma_e} \right]^{\frac{1}{4}} \left[ \frac{J}{10^{18}~ {\rm G/s}}  \right]^{\frac{1}{2}}~.
%\end{equation}
%{\bf where the current scales with the surface value at the pole, and $e^\nu=(1-2M/R)^{1/2}$ is the lapse function accounting for the redshift.} \DV{Fede, check this formula of the lapse function of Schwarschild, I am not sure it is analytical...}
\begin{equation}
T_{\text{bb}} \approx  0.18 ~{\rm keV} \left[ \frac{\Delta r}{1~ {\rm m}}\right]^{\frac{1}{4}} \left[  \frac{10^{17} ~{\rm s}^{-1}}{\sigma_e} \right]^{\frac{1}{4}} \left[ \frac{J}{10^{18}~ {\rm G/s}}  \right]^{\frac{1}{2}}~.
\end{equation}
%\FC{para mi gusto esto no aporta nada, solo es solo una forma de re-escribir (correctamente!) la expresion anterior..}
%\footnote{This factor, for a given $B$, implicitly includes the radiative efficiency and is uncertain by up to one order of magnitude due to the Joule dissipation parameters, $\sigma_e$ and $\Delta r$, related to the specific microphysical properties of the outer envelope/magnetosphere.} 
%$ \left[ \frac{\Delta r}{1 m} \right]^{\frac{1}{4}} \left[ \frac{\eta}{10^3 cm^2 s^{-1}} \right]^{\frac{1}{4}} B_{15}^{\frac{1}{2}} \, R_{10}^{-\frac{1}{2}}$. 

\begin{figure}%[h]
	\begin{center}
		\includegraphics[scale=0.33]{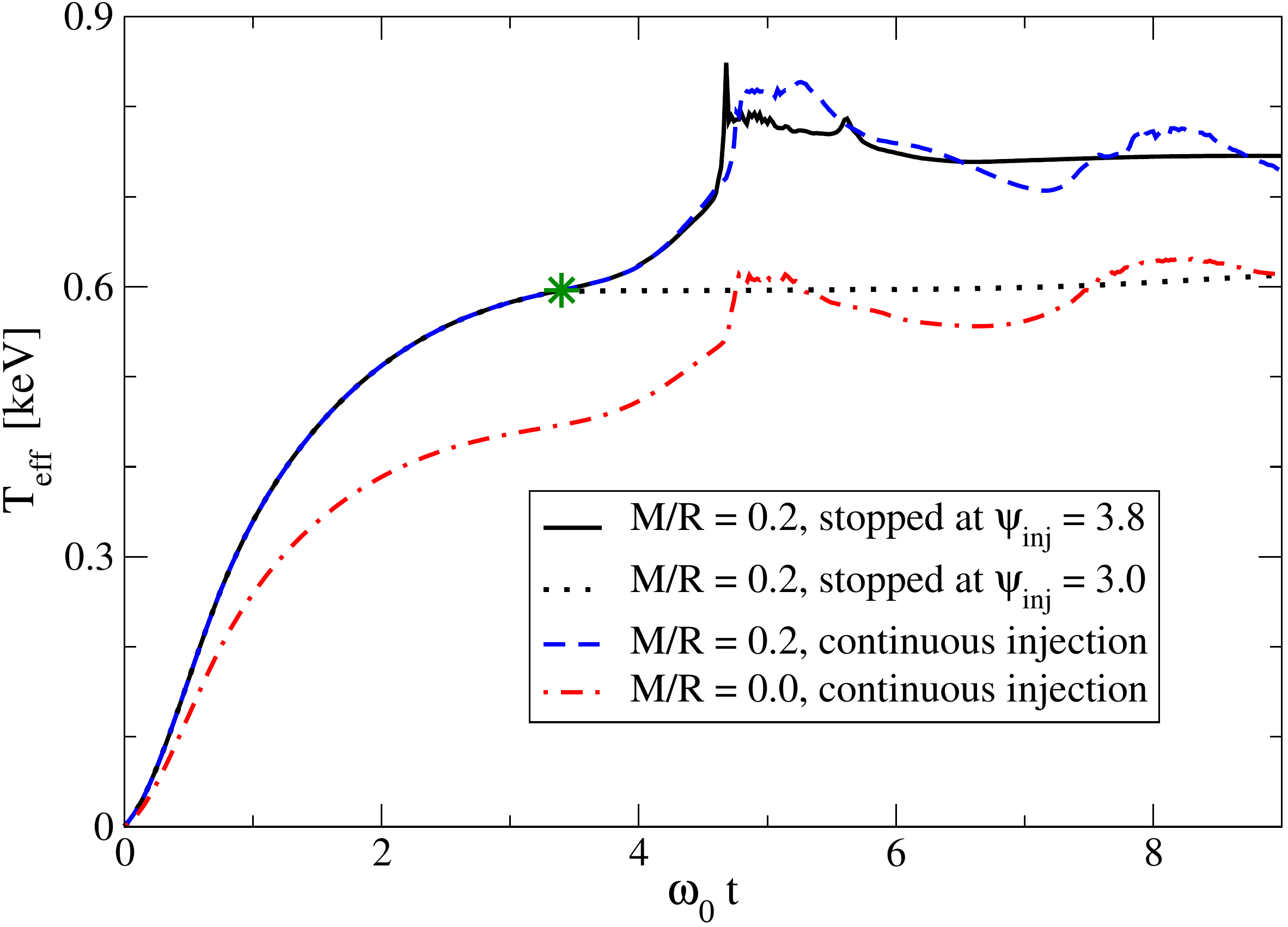}
		\caption{ Effective temperature throughout the evolution of a rotating spot of size $\theta_s = 0.15\pi$ and misalignment $\chi=0.15\pi$. %\DV{Fede, add the symbol at critical temeprature, compute the redshifted temperature and let us see what you prefer. Change all notations and text accordingly}
	The green star symbol points to the critical effective temperature, $T_{\rm crit}\sim 0.6$ keV, prior to the onset of the instability.  
	}
		\label{fig:temperature} 
	\end{center}
\end{figure}

On Fig.~\ref{fig:temperature} we show the effective temperature, defined by the following average over the magnetic spots surface $T_{\rm eff}^4 = < T_{bb}^{4} >$, where $T_{bb}$ is the local surface temperature estimated as described above, assuming the same values as in \cite{akgun2018crust}: $\Delta r =1$ m, and $\sigma_e = 7 \times 10^{16}$ s$^{-1}$. %\DV{Fede, change here accordingly to Temp at infinity depending what you show in the plot}
During the quasi-adiabatic stage, the effective temperature grows with the injected twist (as $T_{\rm eff}\sim \sqrt{\psi_{\rm inj}}$ in the perturbative regime, $\psi_{\rm inj} \lesssim 1$), independently of the particular twist rate employed.
As it is shown in the figure, if the injection is stopped before reaching the critical value, the system remains stable with a constant distribution of the currents. 
Otherwise, the instability is triggered and the temperature quickly increases during the expansion phase. 
During the reconnection event, and soon after, the fireball internal energy and other physical mechanisms (like accelerated particles) will likely provide additional contributions to the X-ray spectra, which can be temporarily dominant. 
However, when the system reaches a stable state (after a few crossing times in our case), the final temperature settles down into an approximately constant value which is always slightly larger than those attained prior to the onset of the instability. Even though this behavior is not intuitive, our interpretation is that the expansion  phase rapidly amplify the magnetospheric currents. While part of these currents are lost during the reconnection, a significant fraction still remain in the final relaxed system, resulting in the observed increase of temperature.

The most extreme case, when a constant twist is injected continuously even after reaching the critical value, shows a similar behavior, with an asymptotic temperature that grows very slowly in the timescale of our simulations.
The plot also illustrates the role of GR effects by comparing the solutions at two different stellar compactness. 
It can be seen that when curvature is neglected ($M = 0$), the effective temperature reduces by apparently $\sim 20\%$. Such increase is due to the combination of the Schwarzschild metric factors entering in the definition of the fields and their derivatives.
However, we note that this is the local temperature in the star. For an observer at infinity the temperature must be red-shifted in the relativistic case, and therefore the observed temperatures would be similar in both cases. %\DV{Check these last sentences} \FC{True!}

\begin{figure}%[t]
  \begin{center}
  \includegraphics[scale=0.33]{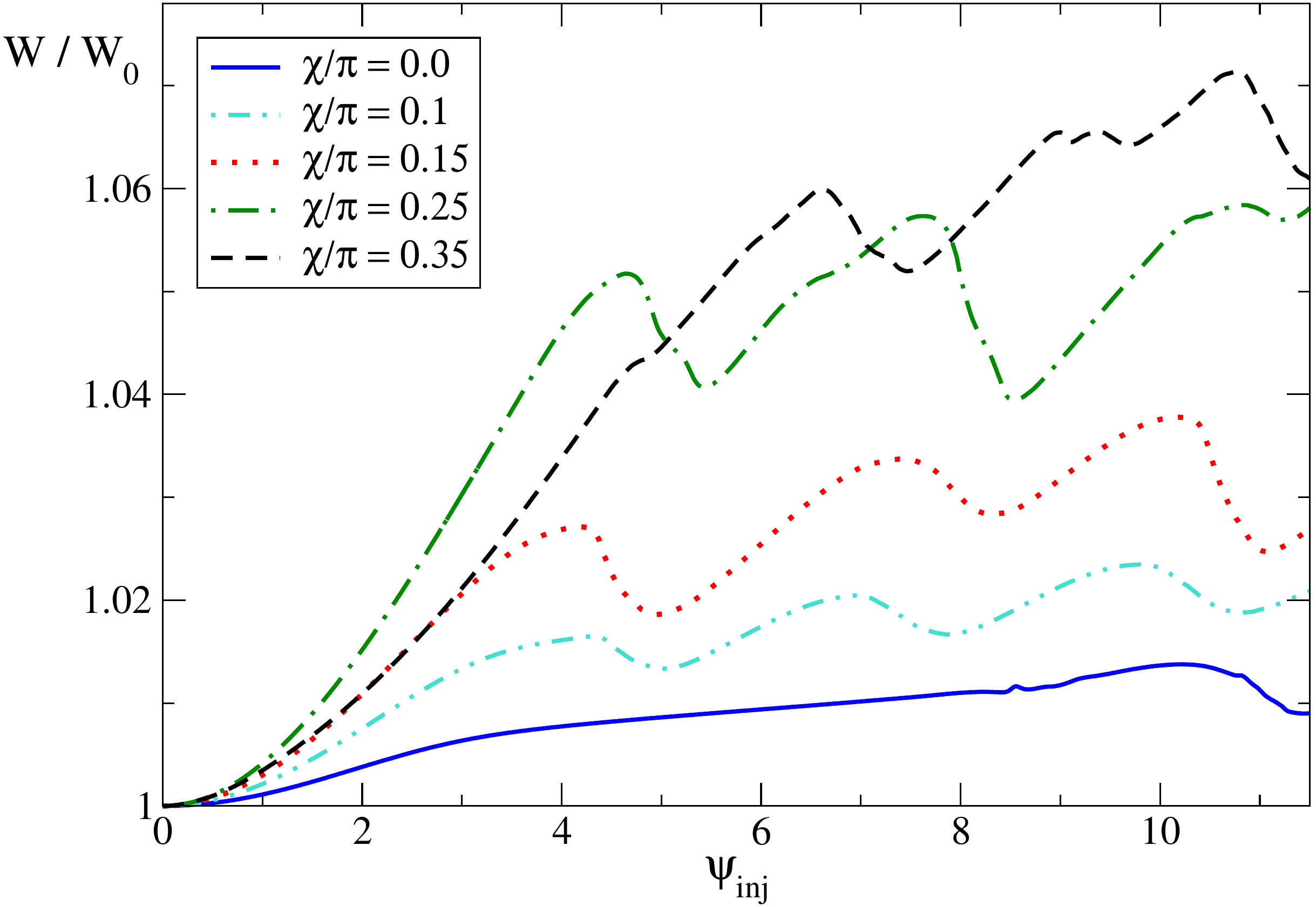}
  \includegraphics[scale=0.33]{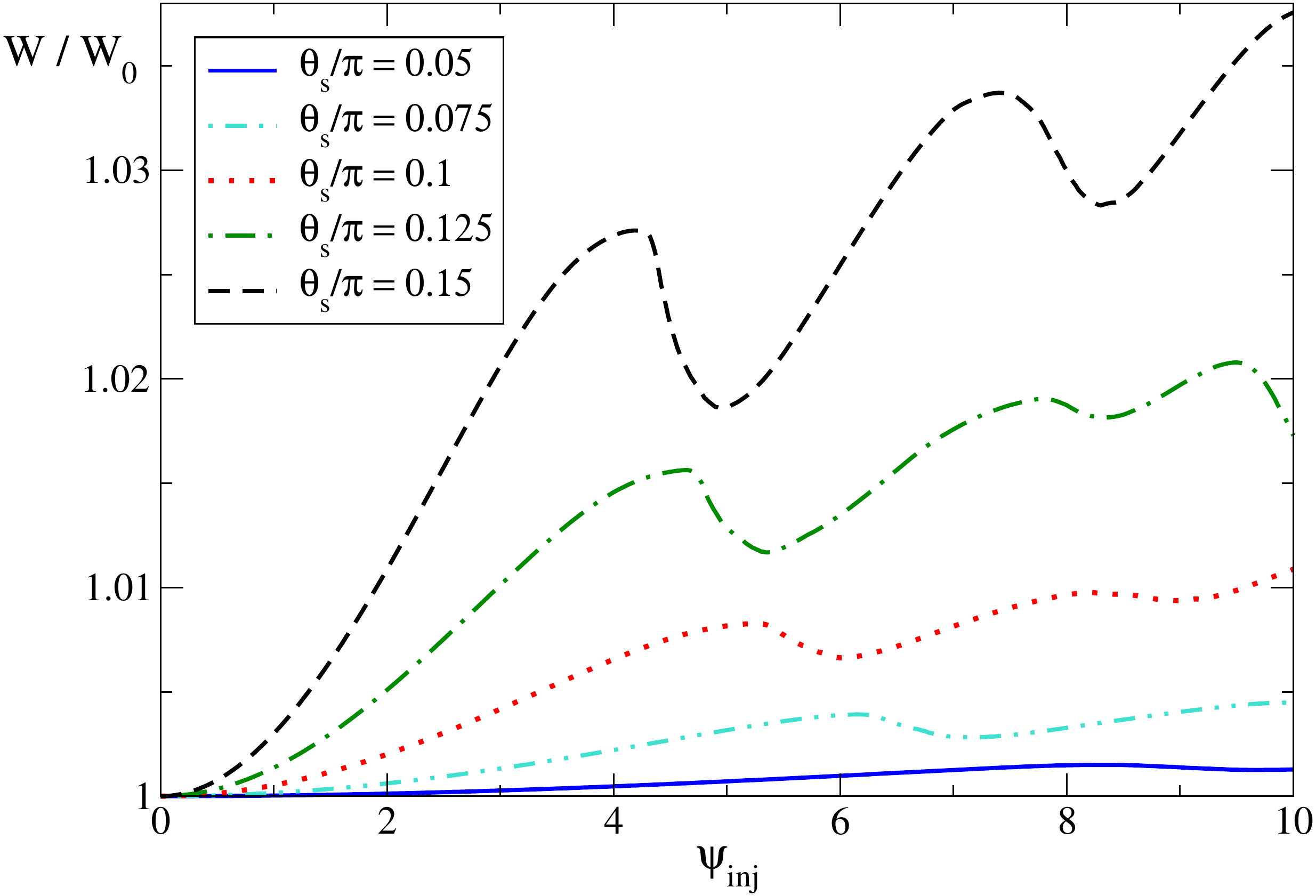}
  \caption{Magnetic energy behavior as a function of the continuously injected twist. 
  Top panel: comparison for different misalignments $\chi$ at spot size $\theta_s = 0.15\pi$. %$\chi/\pi = \left\lbrace 0.0, 0.1, 0.15, 0.25, 0.35 \right\rbrace $.
  Bottom panel: comparison at fixed $\chi = 0.15\pi$ for several spot sizes $\theta_s $. % $\theta_s /\pi = \left\lbrace 0.05, 0.075, 0.1, 0.125, 0.15 \right\rbrace $.
  }
 \label{fig:magnetic_energy} 
 \end{center}
\end{figure}

We also explored different misalignment angles for a fixed spot size of $\theta_s = 0.15\pi$, and different spot sizes for a given angle $\chi = 0.15\pi$.
The results are summarized on Fig.~\ref{fig:magnetic_energy}, where the total magnetic energy in the magnetosphere is displayed as a function of $\psi_{\rm inj}$.
In this sequence of models, we have applied a continuous shear, without turning it off at any moment. Thus, driven by this continuous twist injection, reconnection events occur repeatedly. 
Since each event dissipates or expels only part of the additional magnetic energy, there is an slight increase of the stored energy with time.  
These results provide an estimate of how much energy can be stored and later released depending on the two main parameters of the problem, namely: $\chi$ and $\theta_s$. The available energy stored prior to reconnection increases with both the misalignment angle and with the size of the shearing spots, and is of the order on only a few percent of 
$W_0$. The total twist at which the first reconnection takes place, $\psi_{\rm rec}$, varies between 4 and 11 radians.
However, some cautionary comments are necessary. First, $\psi_{\rm rec}$, depends on numerical resolution and --fundamentally-- on the twisting rate $\omega_o$.
As observed in \cite{parfrey2013}, large values of $\omega_o$ produces a dynamical stabilizing effect which tends to delay the onset of reconnection beyond the critical twist value $\psi_{\rm crt}$. Thus, in general one has that $\psi_{\rm rec} \geq \psi_{\rm crt}$.
We find, nevertheless, that such dynamical stabilization is very prominent in the aligned cases, where the symmetry favors the effect, but it is not that important in the misaligned cases studied here.
%When the axes are misaligned, on the other hand, we observe that $\psi_{rec}$ is much closer to its critical value $\psi_{crt}$, even for the large twisting rates used in this work.
%Furthermore, we notice that these two values approach very quickly, as the misalignment angle is incremented. And hence, we expect the critical twist value $\psi_{crt}$ to increase monotonically with $\chi$,
%as well as the available energy stored prior to reconnection. This energy is also shown to increase with the size of the shearing spots, whereas the reconnection angle $\psi_{rec}$ decreases instead.

\begin{figure}%[h]
	\begin{center}
		\includegraphics[scale=0.33]{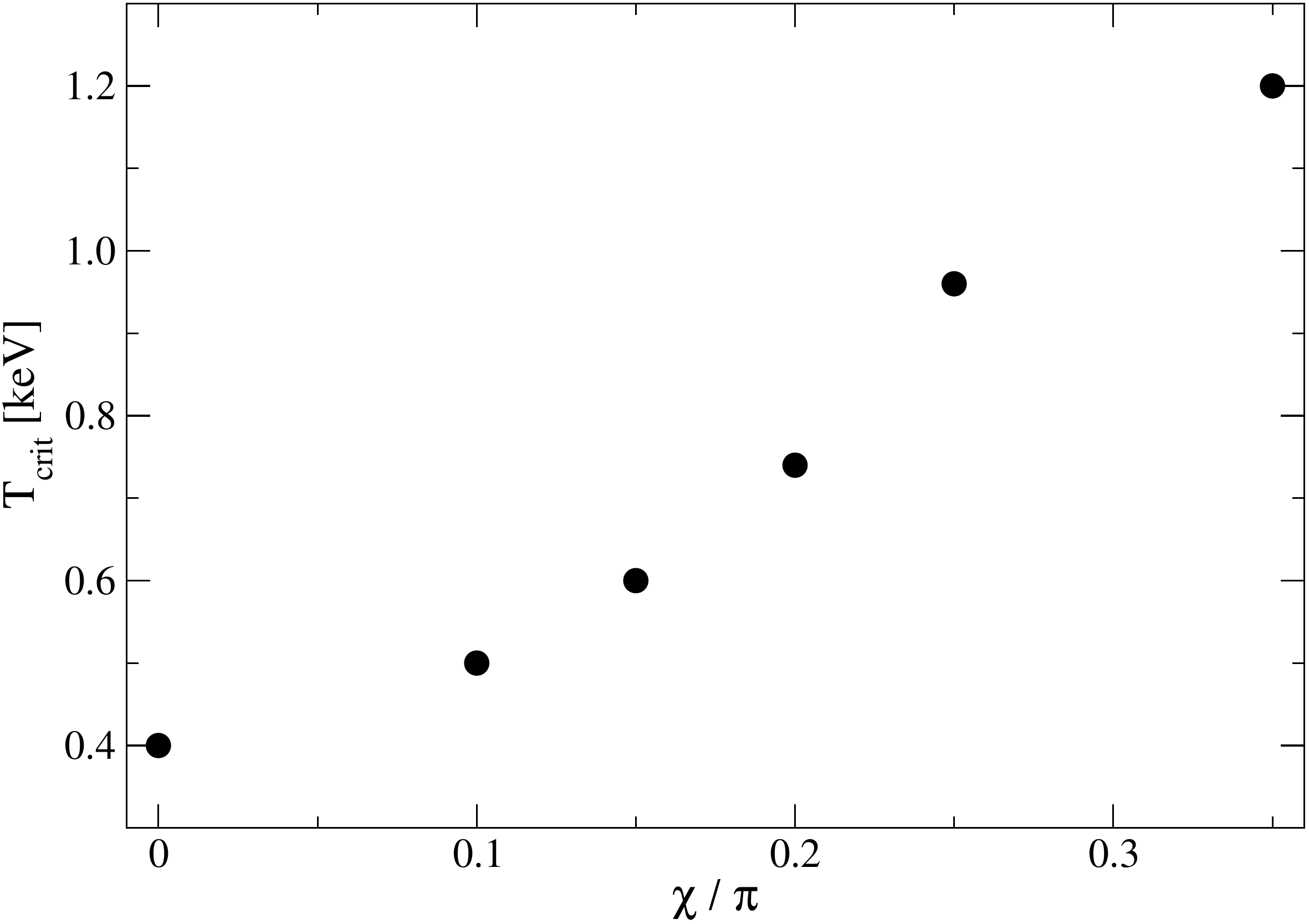}
		\caption{ Estimated effective temperatures near the critical twist value, $T_{\rm crit}$, for different misalignment angles $\chi$. %\jp{I would remove error bars}
		}
		\label{fig:Tcrit} 
	\end{center}
\end{figure}

The maximum temperatures attained near the critical twist value, $T_{\rm crit}$, also increase with the misalignment $\chi$, as shown in Fig.~\ref{fig:Tcrit}: 
the position and geometry of the twisted lines is crucial in determining the heat released at the surface. 
On the other hand, we find that changing the magnetic spot size only affects the global energetics (larger volumes involved), but not the estimated effective temperature. 

%%%%%%%%%%%%%%%%%%%%%%%%%%%%%%%%%%%%%%%%%%%%%%%%%%%%%%%%%%%%%%%%%%%%%%%%%%%%%%%%%%%%%%%%%%%%%%%%%%%%%%%%%%%%%%%%%%%%% CONCLUSIONS

\section{Discussion}
%\noindent{\bf{\em Discussion.}}

In the framework of the standard magnetar model, we have studied the magnetospheric response to the smooth injection of currents from the star interior, supposedly originated 
by the slower internal magnetic field evolution. 
Our results from GR3D simulations show that, starting with a potential model, the magnetosphere gradually stores helicity and energy until a critical twist value is reached, 
after which a reorganization occurs through magnetic reconnection and a new, quasi-steady, magnetospheric configuration is reached. 
The misaligned cases exhibit important differences respect to the axi-symmetric scenarios, in terms of the dependence of the additional magnetic energy stored for a given twist and the temperature of the heated surface on the position of the twisting footprints.

Connecting our results with 
observations pre- and post-outburst  (for the about ten cases where pre-outburst data are available, \cite{coti18}) we can draw the following conclusions. 
First, we find that stable, non-potential configurations can persist after an event, and the estimated temperatures of the magnetic spots
are in line with the high observed values of $0.5$--$1.0$ keV, after the outburst occurs. 
Second, outburst observations show a sudden flux enhancement by up to a few orders of magnitude, accompanied by an increase of the inferred blackbody temperature 
by a factor of a few (even considering the many caveats related with spectral fitting uncertainties). 
We have no precise information about the outburst rise timescale, because no X-ray dedicated observation has ever been lucky enough to catch a magnetar while undergoing an outburst. However, we can place an upper limit at about 1.5 days, which is the closest X-ray observation in quiescent state immediately preceding an outburst \cite{coti18}. 
Since the injection of twist is coupled to a continuous increase of temperature, the fact that the temperature suddenly increases on fast timescales ($<$ one day) indicates that
the last stages of the helicity injection must also be very rapid. This suggests that the outburst trigger is not purely driven by the slow, smooth crustal magnetic evolution (timescales of kyrs at least), because that alone should cause a smaller pre- and post-outburst temperature variation (no more than $\sim 20\%$ in our simulations). 
Instead, the final pre-outburst stage would be compatible with shorter timescale phenomena of sudden magnetic helicity injection (likely triggered, in turn, by the long-term crustal evolution), such as thermo-plastic waves \cite{beloborodov14} or crustal fractures \cite{thompson93,perna11}.
Moreover, for the handful of sources with good-quality data spanning many years, the so-called quiescent state shows no trend of significant spectral/flux evolution 
over years, indicating that stable twisted magnetospheres supported by slow helicity injections maintaining a small twist below the critical value may be very long lived.

A work in preparation is going to further describe the numerical details and explore a wider range of cases, in terms of spot size, location and injection rate. 
It will be important in the future to also consider different initial base configurations and injection distribution, including for instance higher multipoles, 
or allowing the possible interaction between different twisted bundles. 
A connection with the internal evolution of isolated neutron stars will allow to better assess the shape and distribution of very slow twist injection, and the resulting magnetosphere. 
Last, observationally the few magnetars showing repeated outbursts can helpful in partially constraining the location and size of the twisted footprints.

%However, such timescale is dynamically very long, and does not constrain models.
%
%The plasmoid ejection and the energy dissipated at the current sheets (about half of the total loss of energy) could be related with the hard X/$\gamma$-ray emission, similar to the ones seen in the three historical the giant flares, which likely point to a more global reconfiguration compared to the one here simulated (the pulse profiles change after the events). The fact that we do not observe such emission in the standard outbursts point to the fact that either the high energy is strongly beamed, or the efficiency of converting the lost energy into radiation is low. CHECK SOME NUMBERS ABOUT THIS... NOT SURE IF PUT THIS HERE, OR DESERVE A PAPER WITH FRANCESCO, SO HE CAN TELL US SOME MORE PRECISE NUMBERS AND CAVEATS The energetics of the observed short bursts \cite{collazzi15}, on the other hand, are well below the ones inferred from the simulations, thus confirming the picture that these events are likely related with other more localized reconnections, possibly in the high magnetosphere \cite{lyutikov03}. CHECK OTHER REFS ABOUT THE STANDARD PICTURE OF BURSTS/OUTBURSTS. Notice that the study of the global configuration of the magnetosphere is often undertaken under the assumption of force-free models, applied also for the solar corona \cite{mikic1988, biskamp1989,finn1992, cowley1997}. 

%%%%%%%%%%%%%%%%%%%%%%%%%%%%%%%%%%%%%%%%%%%%%%%%%%%%%%%%%%%%%%%%%%%%%%%%%%%%%%%%%%%%%%%%%%%%%%%%%%%%%%%%%%%%%%%%%%%%% AGRADECIMIENTOS

\section*{Acknowledgments}

We acknowledge support from the Spanish Ministry of Economy, Industry and Competitiveness grants AYA2016-80289-P and AYA2017-82089-ERC (AEI/FEDER, UE), and AYA2015-66899-C2-2-P. CP also acknowledges support from the Spanish Ministry of Education and Science through a Ramon y Cajal grant.  
This work used computational resources from \textit{Pirayu Cluster} (supported by the Agencia Santafesina de Ciencia, Tecnolog\'{i}a e Innovaci\'{o}n, Gobierno de la Provincia de Santa Fe, Proyecto AC-00010-18), \textit{Centro de Computaci\'{o}n de Alto Desempe\~{n}o} (CCAD) and \textit{Centro de C\'{o}mputos de Alto Rendimiento} (CeCAR). 
All part of the \textit{Sistema Nacional de Computaci\'{o}n de Alto Desempe\~{n}o},  MinCyT-Argentina. 
The work has been done within the PHAROS COST action CA16214.

%%%%%%%%%%%%%%%%%%%%%%%%%%%%%%%%%%%%%%%%%%%%%%%%%%%%%%%%%%%% BIBLIO

\bibliographystyle{mnras}
\bibliography{FFE}

\end{document}